%% file: v3.tex
\documentclass[11pt,preprintnumbers]{article}
\usepackage[T1]{fontenc}
\usepackage{aas_macros}
\usepackage{mathStyle}
\usepackage{dsfont}
\usepackage[us,12hr]{datetime}
\pdfoutput=1
\usepackage{fancyhdr}
\usepackage{floatrow}
\usepackage{setspace}
\usepackage{algorithm}
\usepackage{algpseudocode}
\usepackage{amsmath,amssymb,amsthm,latexsym,bbm,calc,wasysym,mathtools,empheq, yfonts,upgreek}
\usepackage[english]{babel}
\usepackage[usenames,dvipsnames]{xcolor}
\usepackage{graphicx}
\usepackage[framemethod=tikz]{mdframed}
\usepackage[english]{babel}
\usepackage[utf8]{inputenc}
\usepackage{listings}
\usepackage{amsfonts}

\definecolor{alizarin}{rgb}{0.82, 0.1, 0.26}

\newcommand{\CHANGE}[1]{\textcolor{blue}{{#1}}}
\usepackage{tcolorbox} 
\usepackage{relsize}
\usepackage{caption}
\usepackage{subcaption}
\usepackage{graphicx}
\usepackage[colorinlistoftodos, color=green!40, prependcaption]{todonotes}
\usepackage{braket}
\usepackage{algorithm}
\input{styles} 
\newcommand{\code}[1]{\tcbox[on line,colback=pybackground,colframe=white,boxsep=3pt,left=0pt,right=0pt,top=0pt,bottom=0pt]{\lstinline[language=python,breaklines=true,breakatwhitespace=true]|#1|}}

\newmdenv[innerlinewidth=0.5pt, roundcorner=4pt,linecolor=mauve,innerleftmargin=6pt,
innerrightmargin=6pt,innertopmargin=6pt,innerbottommargin=6pt]{mybox2}

\usepackage[format=plain,
            labelfont={bf,it},
            textfont=it]{caption}

\usetikzlibrary{backgrounds,fit,decorations.pathreplacing,calc}

\def\kp{\ket{\psi}}

\def\a{\alpha}
\usepackage{caption}
\usepackage{subcaption}
\usepackage{graphicx}

\def\Tr{{\rm Tr}}

\def\L{\mathcal{L}}

\def\gO{\tikz\draw[black,fill=black] circle (.3ex);}

\definecolor{rindou1}{rgb}{0.4431,0.2862,0.7960}
\definecolor{rindou2}{rgb}{0.0078,0.1215,0.4392}
\definecolor{lapis}{rgb}{0.0.0470,0.2941,0.5568}
\definecolor{mn}{rgb}{0.15, 0.35, 0.95}

\hypersetup{%
	colorlinks,
	citecolor=magenta,
	filecolor=BrickRed,
	linkcolor=rindou2,
	urlcolor=blue,
	linktocpage=true
}

\begin{document}

\title{\LARGE{\textsc{Sparsity dependence of Krylov state complexity in the SYK model}}} 
\author[\oplus]{\large Raghav G.~Jha,}
\author[\otimes]{\large Ranadeep~Roy} 
\affiliation[\oplus]{\large \vspace{4mm} ~Thomas Jefferson National Accelerator Facility, Newport News, VA 23606, United States}
\affiliation[\otimes]{\large ~The Ohio State University, Columbus, OH 43210, United States}
\emailAdd{raghav.govind.jha@gmail.com}
\emailAdd{ranadeep83@gmail.com} \vspace{5mm}
\abstract{\\{\textsc{Abstract:} 
We study the Krylov state complexity of the Sachdev-Ye-Kitaev (SYK) model for $N \le 28$ Majorana fermions with $q$-body fermion interaction with $q=4,6,8$ for a range of sparse parameter $k$ that controls the number of remaining terms in the original SYK model after sparsification. The critical value of $k$ below which the model ceases to be holographic, denoted $k_c$, has been subject of several recent investigations. Using Krylov complexity as a probe, we find that the peak value of complexity does not change as we increase $k$ beyond $k \ge k_{\text{min}}$ at large temperatures. We argue that this behavior is related to the change in the holographic nature of the Hamiltonian in the sparse SYK-type models such that the model is holographic for all $k \ge k_{\text{min}} \approx k_c$. Our results provide a novel way to determine $k_c$ in SYK-type models.   
}}

\vspace{-20mm} 
\toccontinuoustrue
\vspace{-20mm} 
\maketitle
\textbf{\label{sec:0}\section{Introduction}} 

The holographic conjecture has been a significant advance over the past three decades. It suggests that certain quantum field theories or many-body systems encode gravitational behavior in a specific limit. This is useful because it offers crucial insights into the details of a prospective theory of quantum gravity by understanding quantum many-body systems. The best known example is the SYK model \cite{Sachdev:1992fk, Kitaev:2015, Maldacena:2016hyu} that has been extensively studied using classical~\cite{Kobrin:2020xms, Garcia-Garcia:2023jlu} and quantum computers~\cite{Asaduzzaman:2023wtd, Jha:2024vcw, Araz:2024xkw}. 

In many-body systems with holographic behavior, such as the SYK model, it is always intriguing to probe the extent to which one can simplify the Hamiltonian while still retaining the holographic features. One common measure of this simplification is determined by how many terms we can remove from the original Hamiltonian and still have possible connection to gravitational interpretation. This question has been well studied since the model was proposed, and various observables have been computed, such as the spectral form factor, spectral gap ratio, Lyapunov exponent
\cite{Xu:2020shn, Garcia-Garcia:2020cdo, Garcia-Garcia:2023jlu, Orman:2024mpw} and
the maximum allowed sparsity, or smallest $k$ (defined as the average number of remaining terms in the Hamiltonian normalized by the number of fermions, $N$), to check whether it still retains holographic features. We refer to this as $k_{\text{c}}$ and determining it from the Krylov complexity is the main result of this paper. The reported values of $k_{\text{c}}$ in the literature vary considerably \cite{Xu:2020shn, Orman:2024mpw}. In this paper, we use a well-known observable to characterize the behavior change, which has not been previously considered, and pinpoint the region where the model loses its potential connection to quantum gravity. 

To do this, we make use of the Krylov complexity (or spread complexity) of quantum states under unitary time evolution \cite{Balasubramanian:2022tpr}. This complexity accurately reflects the unique behavior of the model and allows us to calculate the average number of terms required in the Hamiltonian for a holographic description. Complexity, in general, is a broad concept that is studied across different areas. For this work, we will be concerned with the spread complexity of the initial state, which is a measure of how complicated the state becomes under time evolution and how much it spreads in general in Hilbert space. 

The spread complexity has been explored before for both Hermitian and non-Hermitian SYK models
\cite{Balasubramanian:2022tpr, Nandy:2024wwv, Camargo:2023eev, Camargo:2024deu}, so it is not new that we are considering this model. However, as we show, our work is novel in using the saturation of the peak value of Krylov complexity to find the boundary between holographic and non-holographic behavior
in SYK-type models.

Although various notions of complexity have been proposed in recent decades and its relation to holography has also been explored \cite{Susskind:2018pmk, Barbon:2019wsy, Rabinovici:2020ryf}, Krylov complexity has emerged as a promising framework for studying quantum chaos with connections to out-of-time-order correlators (OTOCs) and spectral form factors (SFF) in the last few years. Krylov operator complexity quantifies the growth rate of operators (or how the operator spreads) under the Heisenberg time evolution. It was first shown in~\cite{Parker:2018yvk} that this problem can be modeled as a one-dimensional tight-binding problem, where the position $n$ of a particle in the Krylov chain signifies the complexity of the time-evolved operator. The model incorporates hopping terms between sites, which depend on $\beta_n$, which determine the probability of particle spreading along the chain. The coefficients $\alpha_{n}$ and $\beta_{n}$ are called the Lanczos coefficients, and their growth has been proposed to be related to the chaotic nature of the system as one progresses along the Krylov chain. Following on the proposal in~\cite{Parker:2018yvk}, the authors of~\cite{Balasubramanian:2022tpr} extended the idea to state (or Schr\"{o}dinger) picture where the initial state evolves under a given Hamiltonian and consider the Krylov complexity growth of the state. In chaotic systems, the Krylov complexity of the state tends to rise, reach a peak, and then plateau after some time. This is referred to as the `ramp-peak-slope plateau' and is similar to the `quartic' behavior of spectral form factors (SFF) one encounters in random matrix theory, that is, `slope-dip-ramp-plateau' \cite{Cotler:2016fpe}. 

However, the number of terms in the Hamiltonian for the standard SYK model grows as $\sim N^{4}$ where $N$ is the number of Majorana fermions. This dense nature of the Hamiltonian, is not very suitable for numerical computations, and various works have considered the extent to which the model can be truncated (or sparsified) so that it still retains the holographic features starting with Ref.~\cite{Xu:2020shn}. The basic motivation is to find a model in which the number of terms is proportional to $N$. The coefficient of this scaling, $k$, which results in the number of terms remaining in the Hamiltonian $kN$, where the model is still holographic, has been subject to much investigation~\cite{Xu:2020shn, Orman:2024mpw}. We refer to this as $k_{\text{c}}$ in this paper. 

In this work, we propose additional inference one can draw from the Krylov complexity. It has been well-established that the peak in spread complexity is a characteristic signature of chaotic systems. 
If beyond a certain value of $k$ (which we refer to as $k_{\text{min}}$), the peak value of the spread complexity ceases to increase starting with an initial state (defined later) it means that the system is sufficiently dense and close to the critical sparsity. On the other hand, if for a given $k$ the peak value of the spread complexity has not saturated to its maximal value, the system (described by the Hamiltonian) is not dense enough and thus cannot admit holographic features.  

\begin{figure}
	\centering 
	\includegraphics[width=1.0\textwidth]{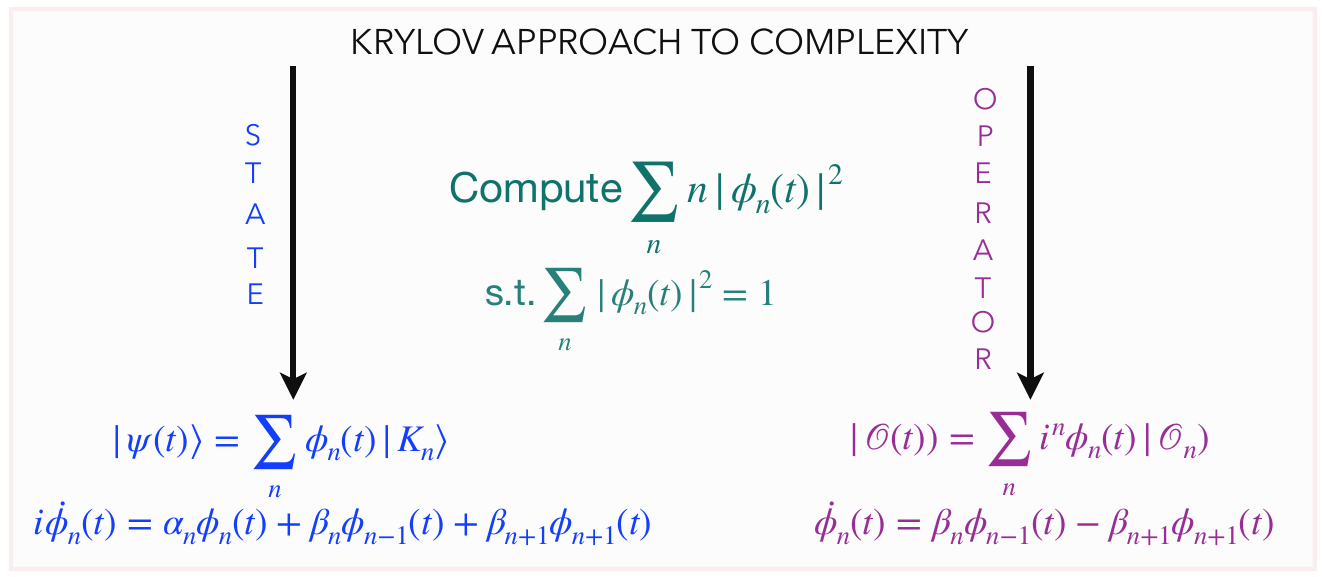}
	\caption{\label{fig:cartoon1}The approaches to Krylov complexity using two pictures (Schr\"{o}dinger and Heisenberg). The operator method was first proposed in \cite{Parker:2018yvk} and the state spread was proposed in \cite{Balasubramanian:2022tpr}. The state corresponding to the operator $\mathcal{O}$ is represented by $\vert \cdot )$ using the standard Choi–Jamiolkowski isomorphism.}  
\end{figure}

We outline and summarize the approach to Krylov complexity in Fig.~\ref{fig:cartoon1} and focus on the state approach to complexity. In Sect.~\ref{sec:sec1} we discuss the sparse and dense SYK models for general $q$ and $N$ Majorana fermions, including the mapping to qubits to construct the Hamiltonian. In Sect.~\ref{sec:sec2}, we discuss the various observables constructed from the Krylov basis, and argue how Krylov complexity can be utilized to detect the critical sparsity for various $N$ and $q$, locate $k_{\text{min}}$ using the peak value of the Krylov complexity as a diagnostic of the change in holographic behavior, and present the results that point to $k_{\text{min}} \approx k_c$ when the temperature is sufficiently large. In Sec.~\ref{sec:TFD}, we consider thermofield double state to explore the temperature dependence as previously considered in various works~\cite{Balasubramanian:2022tpr, Baggioli:2024wbz}. 
We conclude the paper with a summary and future directions and provide additional details in Appendix~\ref{app:sec1}. 

\vspace{5mm}
\noindent \emph{Note added:}
As we were preparing the manuscript for submission, a related preprint was published~\cite{Baggioli:2024wbz} which discusses the results in our paper. Following them, we have also extended our analysis to TFD states which was missing in the former version of the article. We were informed of two more closely related papers  ~\cite{Bhattacharjee:2024yxj,Chapman:2024pdw}. However, our results do not overlap, and our focus is on the saturation of the peak value of the Krylov complexity and the number of terms retained in the sparse SYK model such that it still retains holographic properties. Although the emergence of the peak captures the transition from integrability to chaos has been well studied~\cite{Rabinovici:2022beu}, we propose an additional feature by noting that the saturation of the peak value is closely related to the model changing its holographic property. 

\vspace{5mm} 

\section{\label{sec:sec1}Sparse and dense SYK models}

The model was introduced~\cite{Kitaev:2015} as a $q=4$ all-to-all Hamiltonian. However, $q > 4$ interactions have also been explored~\cite{Maldacena:2016hyu} and the model is also solvable in the large $q$ limit, similar to the $N$ limit. We refer to such versions as the sparse or dense $q$-SYK models. In this work, we will focus on $q = 4, 6, 8$. The Hamiltonian for the general SYK model with $N$-Majorana fermions is:
\begin{equation}\label{eq:SYK_main}
H = \frac{i^{q/2}}{q!} \sum^{N} p_{i_{1}i_{2} \cdots i_{q}} J_{i_{1}i_{2} \cdots i_{q}}~\psi_{i_1} \psi_{i_2} \cdots \psi_{i_q}\,,
\end{equation}
where the factor of $i^{q/2}$ ensures the Hermitian nature of the Hamiltonian and $\psi_i$ are the Majorana fermions ($\psi_i^\dagger=\psi_i$) satisfying the standard anti-commutation relation $\{\psi_i, \psi_j\} = \delta_{ij}$. The probability of retention of any term is determined by $p$. If a uniform random number taken between 0 and 1 is greater than $p$, then $p_{i_{1}i_{2} \cdots i_{q}}$ is set to zero, and if it is smaller than or equal to $p$, then $p_{i_{1}i_{2} \cdots i_{q}}$ is set to 1. The probability $p$ is related to $N$ and $k$ as $p = kN/\binom{N}{q}$ so that the average number of terms retained in the model is $\approx kN$. The smallest value of $k$ where the model is still holographic is referred to as $k_{\text{c}}$ in the literature. The usual dense SYK is obtained if $p=1$ or $k = \binom{N}{q}/N$. The random couplings $J_{i_{1}i_{2}\cdots i_{q}}$ are drawn from a Gaussian distribution with mean and variance given by:
\begin{equation}\label{eq:Jijkl}
\overline{J_{i_{1}i_{2}\cdots i_{q}}}=0\,,\quad \overline{J_{i_{1}i_{2}\cdots i_{q}}^2}=\frac{(q-1)! \mathcal{J}^2}{pN^{q-1}} \,.
\end{equation}
In this work, we will focus on $q = 4, 6, 8$ with $\mathcal{J} = 1$. To construct the Hamiltonian $H$, we first encode the Majorana fermions using the Jordan-Wigner transformation.
The $N$ fermions in \eqref{eq:SYK_main} can be written in terms of the tensor product of $N$/2 Pauli matrices $X,Y,Z$ and the identity matrix $\mathbb{I}$ as:
\begin{align}
\chi_{2k-1} &= \frac{1}{\sqrt{2}} \Big(\prod_{j=1}^{k-1} Z_{j}\Big)X_{k} \mathbb{I}^{\otimes (N -2k)/2}, \nonumber \\
\chi_{2k} &= \frac{1}{\sqrt{2}} \Big(\prod_{j=1}^{k-1} Z_{j}\Big)Y_{k} \mathbb{I}^{\otimes (N -2k)/2},
\end{align}
where the factor of the square root is to ensure the normalization as in Ref.~\cite{Kitaev:2015}. Using this, we can construct the Hamiltonian for sparse and dense SYK models. The size of the Hamiltonian for $N$ is $2^{N/2} \times 2^{N/2}$ i.e., $N/2$-qubit Hamiltonian.

\section{\label{sec:sec2}Krylov observables and limit of sparsity}

In this section, we discuss the observables that we compute using the algorithm described in Appendix~\ref{app:sec1}. The most important observable is $K$-complexity, which has a peculiar structure, and the average value of the peak- and late-time disorder captures the chaotic nature of the Hamiltonian. \textcolor{blue}{} As we will see from our numerical results, there is a dependence on sparsity such that the peak value of Krylov complexity saturates and is independent of $k$ beyond $k \ge k_{\text{min}}$.

\subsection{Spread complexity}
Consider a quantum state $|\psi(t)\rangle$, obtained from an initial state $|\psi_0\rangle$ through a unitary time evolution as: 
\begin{equation}
|\psi(t)\rangle = e^{-iHt} |\psi_0\rangle . \label{eq:unit_transformation}
\end{equation}
One way of characterizing the complexity of the evolved quantum state 
is to find the minimal length in a particular Riemannian geometry~\cite{Dowling:2006tnk}. However, this is not the only possibility and various proposals have been considered including some for quantum field theories \cite{Jefferson:2017sdb}. However, in recent years, inspired by methods in linear algebra, a new definition of complexity has been considered that will be the subject of this work. To compute the complexity of the spread of the initial state (spread complexity), one can expand the evolved state in any choice of orthonormal basis $\mathfrak{B} = \{ |B_n\rangle, n = 0,1,2,\ldots \}$, with $|B_0\rangle = |\psi_0\rangle$. A physically relevant basis is one that minimizes: 
\begin{equation}
C_\mathfrak{B}(t) = \sum_n n  \Big\vert \langle \psi(t) \vert B_n \rangle \Big\vert^{2} , \label{eq:cost_function}
\end{equation}
across all possible choices of $\mathfrak{B}$. We then define the spread complexity as follows:
\begin{equation}
\mathcal{K}(t) = \min_\mathfrak{B} C_\mathfrak{B}(t) , \label{eq:spread_complexity}
\end{equation}
and the basis that minimizes the complexity in \eqref{eq:spread_complexity} is a special basis known as the `Krylov basis' \cite{Balasubramanian:2022tpr}.
Therefore, we will also refer to the spread complexity as $K$-complexity (where $K$ stands for Krylov). One can construct such a basis starting from a given state and Hamiltonian $H$ as detailed in Appendix~\ref{app:sec1} with the Lanczos algorithm~\cite{chen2024lanczosalgorithmmatrixfunctions}. If we take the orthonormal and ordered Krylov basis $\mathfrak{K} = \{ |K_n\rangle, n = 0,1,2,\ldots \}$, with $|K_0\rangle = |\psi_0\rangle$ as 
\( |K_n\rangle \), then  
the $K$-complexity is defined as:
\begin{equation}
C_\mathfrak{K}(t) = \mathcal{K}(t)  = \sum_{n} c_{n} \left| \langle \psi(t) | K_n \rangle \right|^2 = \sum_{n} n\vert \phi_{n}(t)\vert^{2},
\label{eq:def_KC}
\end{equation}
where $n$ is the index of the Lanczos coefficient, $t$ is the evolution time and $\phi_{n}(t) = \langle K_{n} \vert \psi(t) \rangle$ and $\ket{\psi(t)}$ is the evolved state. As summarized in Fig.~\ref{fig:cartoon1}, this notion of complexity is accessible from both operator~\cite{Parker:2018yvk} and state~\cite{Balasubramanian:2022tpr} pictures of quantum dynamics. \CHANGE{In \cite{Balasubramanian:2022tpr}, it was established for chaotic systems spread complexity exhibits a characteristic profile :  ramp, peak, downward slope, and plateau. At the same time, it is understood that sparse SYK models exhibit a maximally chaotic gravitational sector at low temperatures provided it is dense enough \cite{Xu:2020shn}. This suggests that if we can identify the value of $k$ at which the spread complexity ceases to increase, it would mean that the system has become maximally chaotic\footnote{We refer the reader to Ref.~\cite{Baggioli:2024wbz} for the temperature dependence of Krylov complexity in sparse SYK models and Sec.~\ref{sec:TFD} of this paper.}} We start with a uniform superposition state and compute the complexity to late times $\mathcal{J}t/\mathcal{D}$ where $\mathcal{J}$ is the disorder strength as defined in ~\eqref{eq:Jijkl} and $\mathcal{D}$ is the dimension of both the Hilbert and the Krylov space. We vary the average number of terms in the Hamiltonian by changing $k$ for fixed $N$ and compute the Krylov complexity. As we increase $k$, we find that the peak value of complexity increases, becoming independent of $k$ beyond a certain value of $k$ i.e., $k_{\text{min}}$. The determination of $k_{\text{min}}$ is as follows. We compute $\mathcal{K}(t)$ for different $k$ i.e., $1 \ll k \ll N^{q-1}/q!$ for the $q$-SYK model. We refer to models with $k \sim N^{q-1}/q!$ as \emph{dense} SYK models.\footnote{This is because $k = \binom{N}{q}/N \sim \frac{N^{q-1}}{q!}$ for large $N$} We compute the peak value of $\mathcal{K}(t)$ and denote it by $\widetilde{K}_{( \cdot )}$ where the subscript refers to the number of terms in the Hamiltonian. We take the initial state in this work to be a uniform superposition (pure) state, i.e. $\ket{\psi_0} = (\mathbb{H}\ket{0})^{\otimes N/2}$ where $\mathbb{H}$ is the Hadamard gate. We also consider the thermofield double (TFD) state in Sec.~\ref{sec:TFD} to include the temperature dependence. We then compute the difference of maximum complexity in the dense model and Hamiltonian with $kN$ terms and minimize it over $k$ i.e.,
\begin{align}
    k_{\text{min}} = \text{min}\quad
    k \in \Bigg(\frac{1}{q}, \binom{N}{q}\Bigg),&\\
    \text{s.t.} \quad
    \frac{\Big \vert \widetilde{K}_{\text{dense}} - \widetilde{K}_{kN} \Big \vert}{\widetilde{K}_{\text{dense}}}  \le \epsilon 
    \label{eq:kcrit_estimate}
\end{align}
where $\epsilon$ is the allowed error in our estimate which we set to $\epsilon = 0.02$ due to the finite precision due to the number of samples we considered in this work.   

\subsection{K-entropy and state localization}

In addition to $K$-complexity, one can also define the entropy by noting that $|\phi_{n}(t)|^2$ which describes the absolute value of $\phi$ at index $n$ of the Krylov chain at time $t$ sums up to one~\cite{Barbon:2019wsy}. It is defined as:
\begin{equation}
    \mathcal{S}_{K}(t) = -\sum_{n = 0}^{\mathcal{D}-1} |\phi_{n}(t)|^2 \log |\phi_{n}(t)|^2.
\end{equation}
    
Another useful observable to understand the localization of the state in the Krylov basis is to dynamically assess the strength of wave function localization in the Krylov basis, we can use the extension of inverse participation ratio \cite{Wegner1980} in this basis referred to as Krylov inverse participation ratio \cite{Bhattacharya:2024hto}, $R$, defined as: 
\begin{equation}
    R(t) = \sum_{i=0}^{\mathcal{D}-1} |\phi_{n}(t)|^4.
\end{equation}
Higher values of $R$ indicate a more localized state, with an upper bound of 1. The minimum value is $1/\mathcal{D}$ when completely delocalized, where $\mathcal{D}$ is the dimension of the Krylov space and also the Hilbert space i.e., $\mathcal{D} = 2^{N/2}$.
Our results indicate that (exponential of) Krylov entropy can also be used to determine critical sparsity.

\subsection{Results}

\begin{figure}
    \centering
    \includegraphics[width=0.95\linewidth]{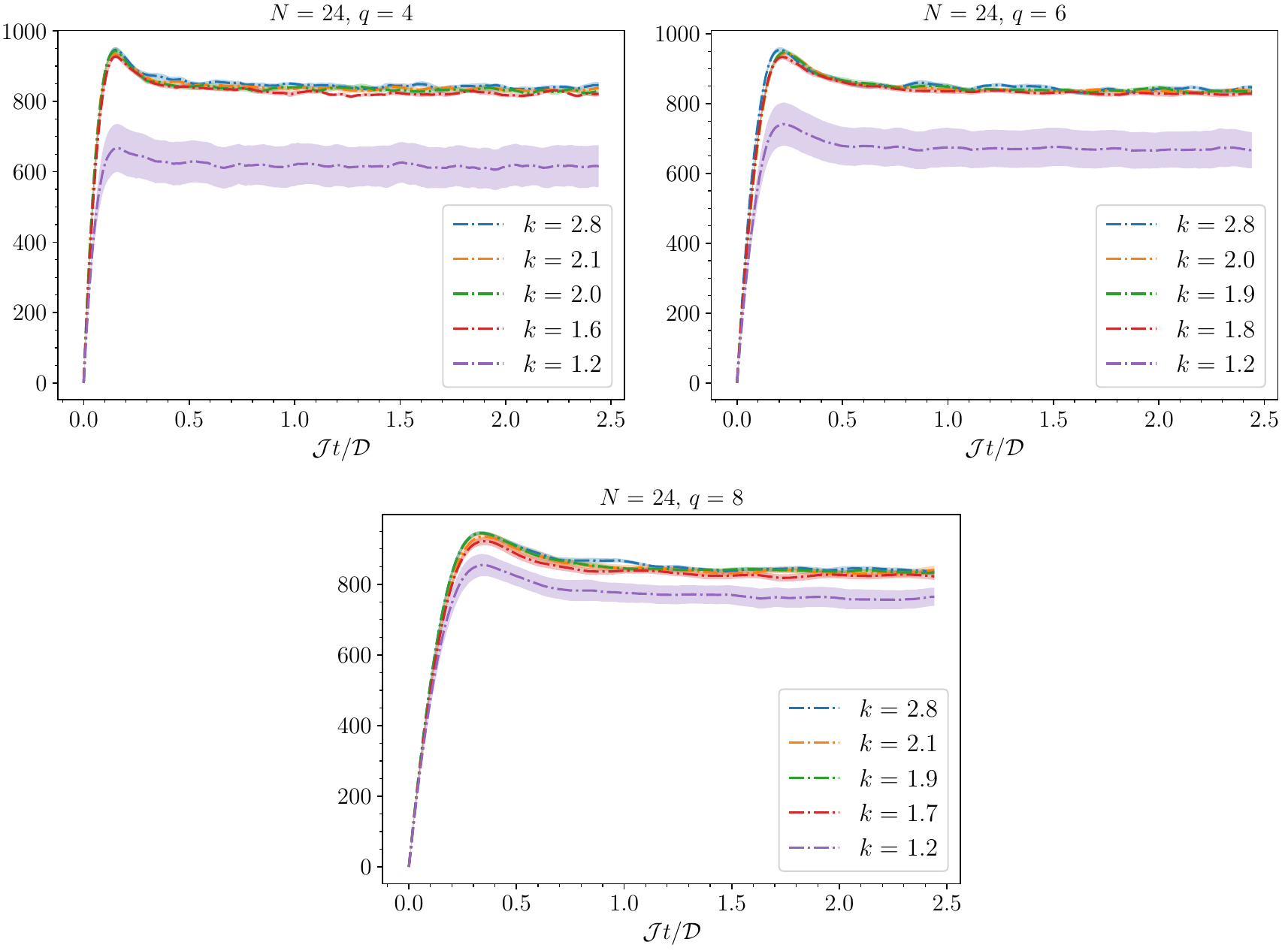}
    \caption{\label{fig:N24main}The time evolution of complexity $\mathcal{K}$ for $N=24$ with $q=4,6,8$. The late-time value of the peak of Krylov complexity increases with $k$ and becomes independent for $k > k_{\text{min}}$. We find the standard `ramp-peak-slope-plateau' behavior. The shaded region denotes the error due to the disorder averaging over multiple instances of the model.}
\end{figure}

\begin{figure}
    \centering
    \includegraphics[width=0.95\linewidth]{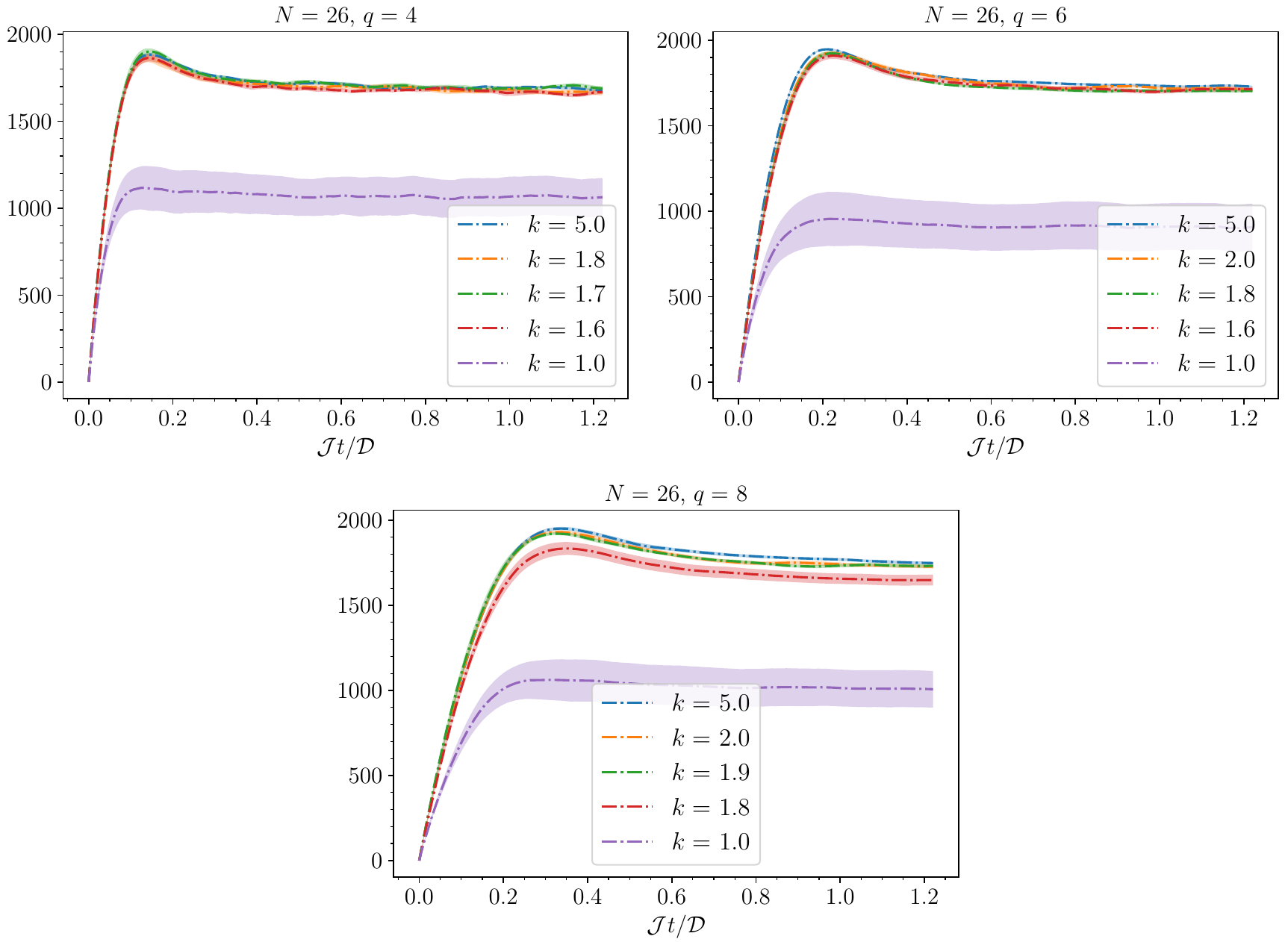}
    \caption{\label{fig:N26main}The evolution of complexity $\mathcal{K}$ for $N=26$ with $q=4,6,8$. The late-time value of complexity and the peak increase with $k$ and become independent for $k > k_{\text{min}}$. The shaded region denotes the error due to the disorder averaging over multiple instances of the model.}
\end{figure}

\begin{figure}
    \centering
    \includegraphics[width=0.95\linewidth]{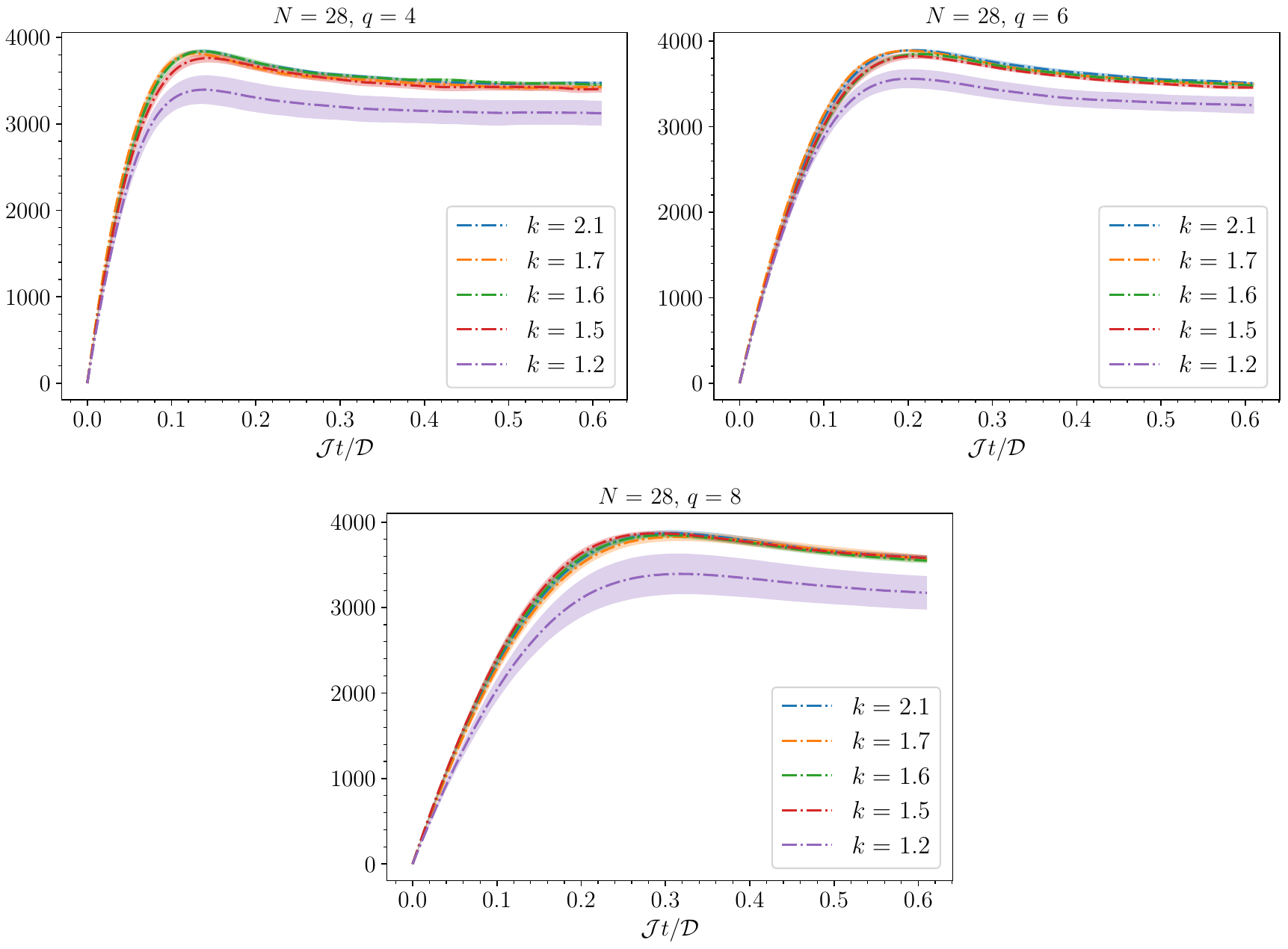}
    \caption{\label{fig:N28main}The evolution of complexity $\mathcal{K}$ for $N=28$ with $q=4,6,8$. The shaded region denotes the error due to the disorder averaging over multiple instances of the model. Our results show that the model is holographic (with the errors due to a small number of samples) for $k_{\text{min}} = 1.5$ which corresponds on average to 42 remaining terms in the Hamiltonian.}
\end{figure}

\begin{figure}
    \centering
    \includegraphics[width=0.95\linewidth]{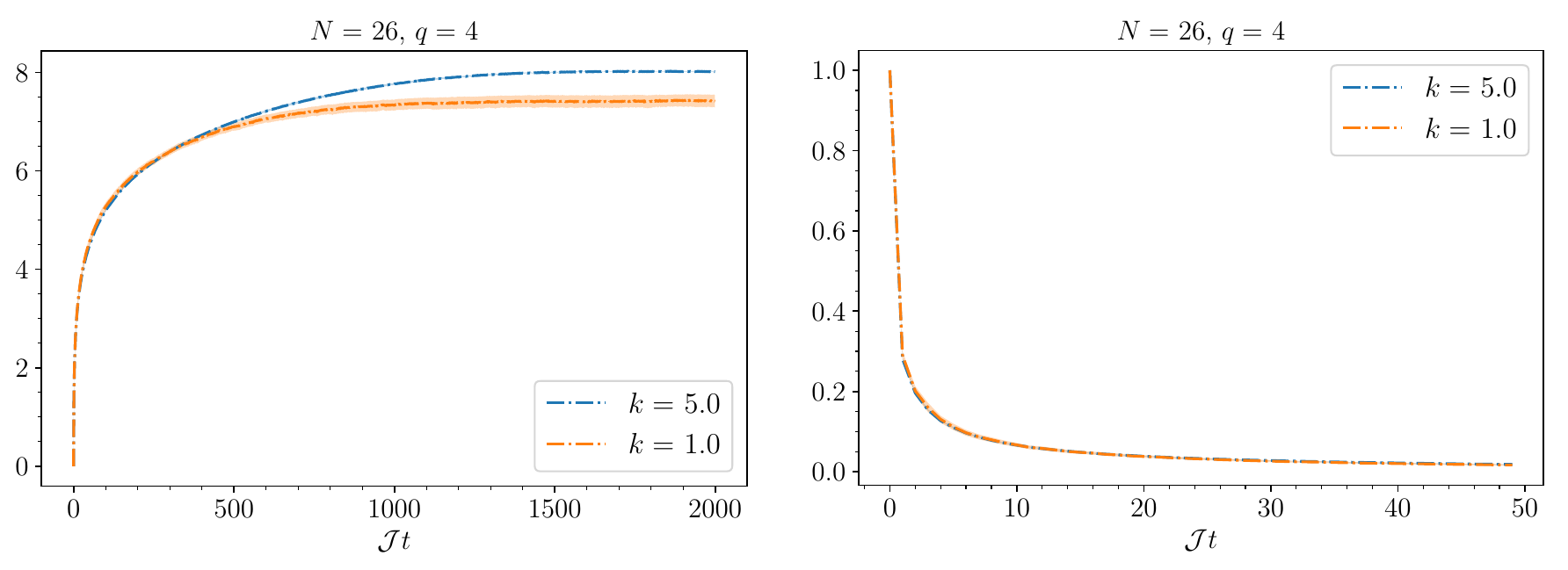}
    \caption{Left: The evolution of the K-entropy for $N=26, q=4$. The entropy for both $k$ increases similarly until $\mathcal{J}t \approx 500$ and then depending on the sparsity, the rate of increase and plateau value are different. Right: The evolution of inverse participation ratio (IPR) for $N=26, q=4$ for $k = 1, 5$.}
    \label{fig:N26plot}
\end{figure}

Using the peak value of the complexity $\mathcal{K}$, we determine $k_{\text{min}}$ for various $N$ and $q$ as shown in Table~\ref{tab:data}. As representative examples of how the various K-observable behave at different times, we show the Krylov spread complexity for $N = 24, 26, 28$ for different $q$ in Figs.~\ref{fig:N24main},~\ref{fig:N26main}, and~\ref{fig:N28main} respectively. The critical sparsity from our computations is given in Table~\ref{tab:data} for $N = 16, 20, 22, 24, 26, 28$ and $q = 4,6,8$. Our results show a negligible dependence of $k_{\text{min}}$ on $q$ with increasing $N$. We also find that $k_{\text{min}}$ decreases as we increase $N$ 
and is independent of $N$ for large $N$. Our results suggest that for any $q$ at large $N$, keeping $3N/2$ terms on average in the Hamiltonian results in a model with the same holographic properties as the dense SYK model. 
For the $K$-entropy, we found that the initial growth follows the exponential growth of the $K$-complexity and the time scale associated with the plateau of the entropy is close to the peak of the spread complexity and is around $ \mathcal{J}t \sim 1250$. The results are shown in~Fig.~\ref{fig:N26plot}.
The results for the inverse participation ratio for $N=26$ and $q=4$ is shown in Fig.~\ref{fig:N26plot}. Starting with $t=0$, we find that the model delocalizes quickly and independently of the sparsity $k$. 


\begin{table}[]
\centering
\renewcommand{\arraystretch}{1.1}
\setlength{\tabcolsep}{6pt}
\begin{tabular}{|c|c|c|c|}
\hline 
$N$ & $q=4$ & $q=6$ & $q=8$ \\ [1.ex]
\hline  \hline
16 & 2.6 & 2.3 & 2.2 \\ [1.ex]
\hline 
20 & 2.5 & 2.2 & 2.1 \\ [1.ex]
\hline 
22 & 2.2 & 2.1 & 2.0 \\ [1.ex]
\hline 
24 & 2.0 & 1.9 & 1.9 \\ [1.ex]
\hline 
26 & 1.7 & 1.5 & 1.9 \\ [1.ex]
\hline  
28 & 1.5 & 1.5 & 1.5 \\ [1.ex]
\hline 
\end{tabular}
\caption{The obtained values of $k_{\text{min}}$ for different $N$ and $q$ determined using~\eqref{eq:kcrit_estimate} using the superposition state. The number of terms in the sparse Hamiltonian is $kN$ on average while the number of terms in the dense Hamiltonian is
$\binom{N}{q}$ which grows like $N^q/q!$. The computed values are disorder averages over at least ten instances of the model.} 
\label{tab:data}
\end{table}

\section{\label{sec:TFD}TFD state and temperature dependence}

\begin{figure}
    \centering
    \includegraphics[width=0.95\linewidth]{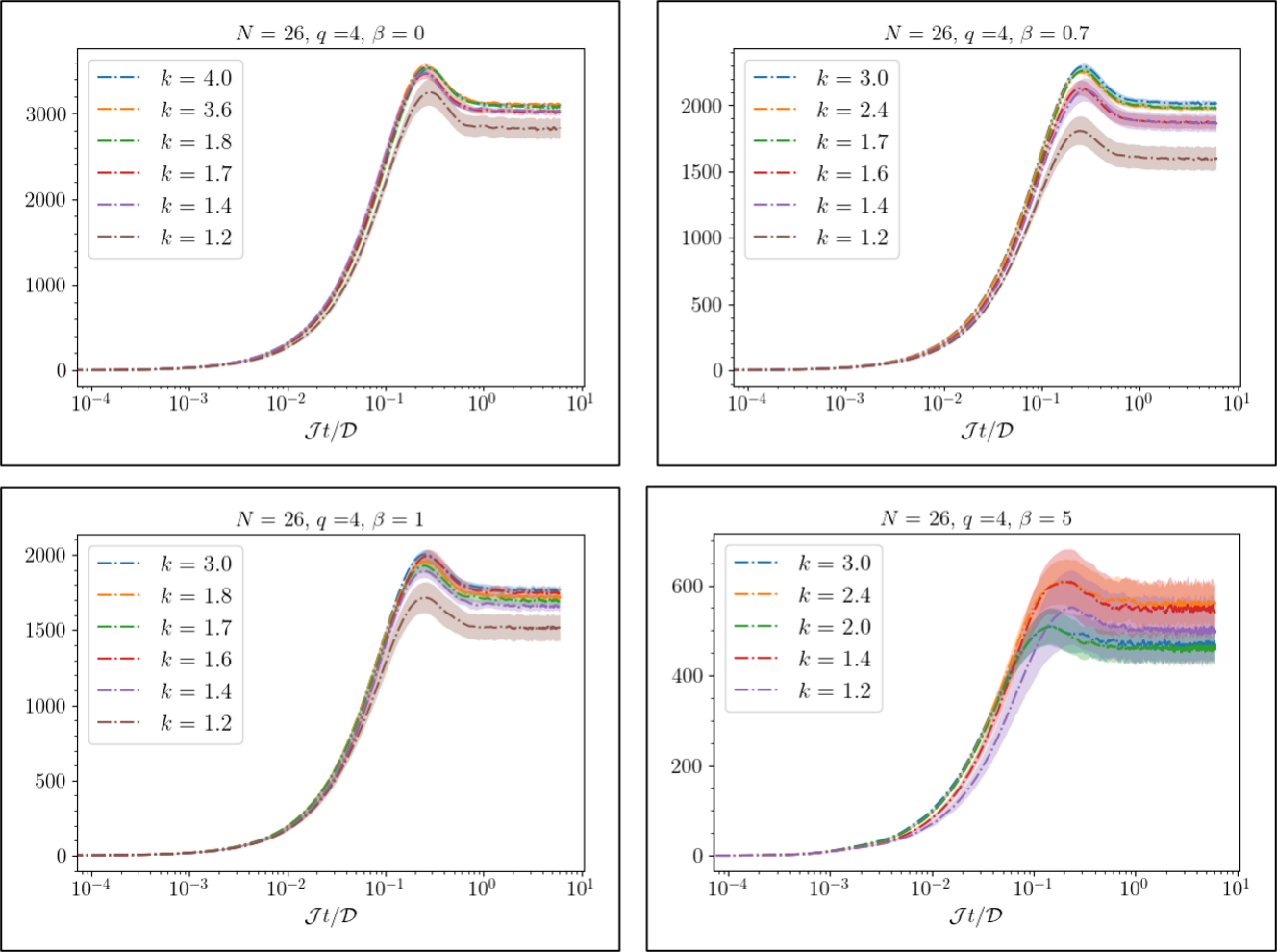}
    \caption{\label{fig:N26tfd}Evolution of complexity $\mathcal{K}$ for $N=26$ with $q=4$ at $\beta=0,0.6,1, 5$ As $\beta$ is increased from $\beta = 0$, the peak saturation which determines $k_{\text{min}}$ can occur rather quickly and is lower than the critical sparsity of the model.
    } 
\end{figure}

Finally, we compute the spread complexity with the thermofield double (TFD) state as the initial state and explore the temperature dependence as shown in Fig.~\ref{fig:N26tfd}. The TFD state is pure state determined up to phase in the doubled Hilbert space and defined as: 
\begin{equation}
    \ket{\text{TFD}} = \sum_n \frac{1}{\sqrt{Z_\beta}} e^{\frac{-\beta E_n}{2}}\ket{n} \ket{n}
\end{equation}
where $Z(\beta)$ is the thermal partition function of the original system and $\{ \ket{n}\}$ denotes the set of eigenstates of the original Hamiltonian. We take the sub-systems to be the same disordered Hamiltonians i.e., $H=H_L=H_R$. The time evolution of this TFD state~\cite{Balasubramanian:2022tpr} governed by the Hamiltonian:
\begin{equation}
    \mathbb{H} = (1/2)(H_L \otimes \mathbb{I}  + \mathds{1}\otimes H_R),  
\end{equation}
and given by:
\begin{equation}
    |\psi(t)\rangle = e^{-i t \mathbb{H}} \ket{\text{TFD}},
\end{equation}
is contained within the space spanned by 
$\{ \ket{n}, \ket{n}\}$. The algorithm for computing 
Krylov complexity works in the subspace where the maximum dimension of the explored Hilbert space is 
$2^{n}$ even though the TFD state is of dimension $2^{2n}$. 
For $N=26$ and $q=4$, we find that $k_c \approx 1.7$ in the high-temperature limit ($\beta \leq 0.6$) with both the superposition state and the TFD state for $q=4$. However, as $\beta$ increases, it appears that one cannot draw a similar conclusion for $q=4$. This means that for the low-temperature (large $\beta$) regime where quantum chaos is expected to saturate the chaos bound for the SYK model, we cannot relate\footnote{We thank the referees for emphasizing this point} the saturation of the peak of Krylov complexity directly to the critical sparsity with the numerical data available to us. We will address this at larger $N$ and explore $q$ dependence in a future work.
\section{Summary}

We computed the critical sparsity (number of terms remaining) in the dense SYK model Hamiltonian such that it still retains holographic features from Krylov complexity. To do this, at large temperatures, we used the saturation of the peak value of the Krylov state spread complexity as a probe. We find that for $k \gtrsim k_{\text{min}}$, the peak value is independent of $k$ up to the dense limit~$\sim N^{q-1}/q!$. 

We also studied the dependence on $q$ and found that $k_{\text{min}}$ has very mild or no dependence on the $q$-body interaction at sufficiently large $N$. Our results and those obtained recently in Ref.~\cite{Nandy:2024wwv} appear to imply that the critical sparsity for both the Hermitian and non-Hermitian SYK models is approximately the same.

Several previous works have investigated the calculation of the spectral form factor, spectral gap ratio, or Lyapunov exponent as a measure of locating the limit of sparsification in the SYK Hamiltonian. The approach we have taken is a fundamentally new way of determining this behavior change and is computationally cheaper and easier to implement than computing the entire spectrum of the Hamiltonian as needed for the spectral gap ratio \cite{Giraud:2020mmb} or complicated four-point our-of-time correlators. Our estimate of $k_{\text{min}}$ for various $q$ also clarifies the existing estimates in the literature \cite{Xu:2020shn, Orman:2024mpw} which differ over a wide range. Obtaining new results for $q=6,8$, we find that our results for $q=4$ are consistent with $k_c<4$ found in \cite{Xu:2020shn} and $k_c \gtrsim 1$ \cite{Garcia-Garcia:2020cdo}.

The peak value of Krylov complexity is an interesting method for tracking when the Hamiltonian becomes dense enough to admit holographic properties. It would be interesting to understand the fine-grained properties of complexity and its bulk interpretation, and some work in this direction has already been done~\cite{Rabinovici:2022beu}. In addition, with recent interest in quantum computing methods, a more ambitious goal would be to study the spread complexity using Krylov methods adapted to quantum hardware~\cite{Yoshioka:2024lle}. However, currently, the task appears to be challenging due to the requirement of a large Krylov space dimension, the need to maintain orthogonality, and resource requirements for the quantum circuit.

\section*{\centering Acknowledgments}
This material is based on work supported by the US Department of Energy, Office of Science, Contract No. DE-AC05-06OR23177, under which Jefferson Science Associates, LLC operates Jefferson Lab. The research was also supported by the U.S. Department of Energy, Office of Science, National Quantum Information Science Research Centers, Co-design Center for Quantum Advantage under contract number DE-SC0012704.

\section*{\centering Code Availability}
The code used for this work written in \code{Python} can be obtained from Ref.~\cite{code}.  

\appendix

\section{\label{app:sec1}Review of Krylov approach to complexity and numerical details}

The use of Krylov subspace techniques finds multiple applications in various areas of physics, such as in lattice field theory for the computation of some eigenvalues of the fermion operator or in the implementation of the conjugate gradient. In addition, Krylov techniques are also popular for the time evolution of quantum systems, for example, see \cite{Garcia-Garcia:2023jlu}. They have also been utilized for quantum reservoir computing~\cite{Domingo:2023kjr}. However, the use of Krylov subspace methods to define a measure of complexity was first proposed only in Ref.~\cite{Parker:2018yvk}.

\subsection{Operator growth} 
The Heisenberg equation of operator growth ($\hbar=1$) is given by:
\begin{equation}
\partial_{t} O(t) = i[H, O],
\label{eq:heisen1}
\end{equation}
The solution to this is the time-dependence 
of growth of operator:
\begin{equation}
\mathcal{O}(t) = e^{iHt} \mathcal{O} e^{-iHt}, 
\label{eq:heisen2}
\end{equation}
which using the BCH formula can be written as:
\begin{align}
\mathcal{O}(t) &= e^{i \mathcal{L} t} \mathcal{O}  \nonumber \\
& = \sum_{n=0}^{\infty} \frac{(it)^n}{n!}\mathcal{L}_{H}^{n}\mathcal{O}  \nonumber \\
& = \mathcal{O} + it [H,\mathcal{O}] + \frac{(it)^2}{2!} [H,[H,\mathcal{O}]]  + \cdots 
\end{align}
We can consider this evolution as the Schr\"{o}dinger’s time evolution where $\mathcal{O}(t)$ plays the role of operator’s wave function, and the Liouvillian $\mathcal{L}$ as the Hamiltonian. One can then 
use a smooth version of `ket' i.e., $\vert \mathcal{O})$ (standard notation) to denote the Hilbert space vector corresponding to the operator 
$\mathcal{O}$ as:
\begin{equation}
\mathcal{O} \equiv \vert \widetilde{\mathcal{O}_0}),~\cdots,~\mathcal{L}^{n} \mathcal{O} \equiv \vert \widetilde{\mathcal{O}_n}).
\end{equation}
normalized to 1 using Hilbert-Schmidt inner product i.e., $(A|B) = \Tr[A^{\dagger}B]$. Several other definitions of inner products such as Wightman product have also been used in the literature.
Starting with $\vert\widetilde{\mathcal{O}_0})$, 
we can use a version of the Gram–Schmidt orthogonalization 
procedure to obtain the Krylov basis. The Krylov space associated with the operator can be defined as the span of nested commutators of the Hamiltonian with the operator:
\begin{equation}
\mathcal{H}_\mathcal{O} = \text{span} \Big\{\L^{n}\mathcal{O}\Big\}_{n=0}^{\mathcal{D}-1} = \text{span} \Big\{\mathcal{O}, [H, \mathcal{O}], [H, [H, \mathcal{O}]], \cdots \Big\}
\end{equation}
The iterative procedure gives us the coefficients $\beta$ which can then be used to obtain $\phi_{n}(t)$ as a solution to the first-order differential equation:
\begin{equation}
\dot{\phi}_{n}  = -\beta_{n+1}\phi_{n+1} - \beta_{n}\phi_{n-1}. 
\end{equation}
Note that compared to spread complexity, we need only $\beta$ coefficients and not $\alpha$. In the case of open quantum systems, the Liouvillian is replaced by $\mathcal{L}$ which is the Lindblad or GKSL (Gorini-Kossakowski-Sudarshan-Lindblad\footnote{The paper by Lindblad was received on 07 April 1975 and published in June 1976 while the one by GKS was received on 19 March 1975 and published in May 1976.}) operator \cite{Gorini1976,Lindblad1976}
and acts as: 
\begin{equation}
 \mathcal{L}\mathcal{O} = [H, \mathcal{O}] -i
 \sum_{k} \Bigg[L_{k}\mathcal{O} L_{k}^{\dagger} - \frac{1}{2} \Big\{ L_{k}^{\dagger}L_{k}, \mathcal{O} \Big \} \Bigg]
\end{equation}
where $L$ represents the jump operators (also called Lindblad operators) introduced in \cite{Lindblad1976}. The simplest jump operator in the SYK model is:
\begin{equation}
    L_i \propto \psi_i, \quad i = 1, 2, \ldots, N.
\end{equation}
where $N$ is the number of Majorana fermions. We refer to 
\begin{equation}
D(t)\gO =  \sum_{k} \Bigg[L_{k}\gO L_{k}^{\dagger} - \frac{1}{2} \Big\{L_{k}^{\dagger}L_{k}, \gO \Big \} \Bigg]    
\end{equation}
as dissipator. In this work, we only consider a closed quantum many-body system. For discussion related to non-Hermitian SYK systems using Arnoldi or bi-Lanczos algorithm, we refer the reader to Ref.~\cite{Bhattacharjee:2022lzy}.

\subsection{State approach}
In the unitary (closed) evolution, suppose that
we start with a quantum state $\kp$ and a time-independent Hamiltonian $H$ which corresponds to the sparse or dense SYK model. The unitary time evolution of the state is determined by the time-dependent Schr\"{o}dinger equation
\begin{equation}
i\partial_t \left|\psi(t)\right\rangle = H\left|\psi(t)\right\rangle.
\label{eq:SE1}
\end{equation}
The solution to~\eqref{eq:SE1} is $\left|\psi(t)\right\rangle = e^{-iHt}\left|\psi(0)\right\rangle$ and is equivalent to solving $\vec{x} = A^{-1} \vec{b}$. The Krylov subspace method to solve for $\vec{x}$ is a well-known method. The Krylov basis has dimension $\mathcal{D}$ which is often smaller than the Hilbert space dimension, so in such cases the usual Krylov basis does not span the full Hilbert space, and finding the solution is
more efficient. 
This is due to the well-known result that given any non-singular $A$ of shape $d \times d$, then the solution vector $\vec{x}$ is contained in the Krylov basis i.e., 
\begin{equation}
 \vec{x} = A^{-1}b \in \text{span}\{b, Ab, A^2b, \cdots, A^{n-1}b\}.   
\end{equation}
where $n \le d$. To prove this, we can use the Cayley-Hamilton theorem. Let $\chi(t) $ be a polynomial function of $t$ defined as $\chi(t) = \det(t \mathbb{I} - A)$ which is the characteristic polynomial of $A$, i.e., $\chi(A) = 0$. 
The Cayley-Hamilton Theorem states that if $A$ is a matrix of size $d \times d$ then, 
\begin{equation}
 A^n + \gamma_1 A^{n-1} + \cdots + \gamma_n  \mathbb{I} = 0.   
\end{equation}
Solving for the identity matrix by rearranging terms and dividing by $\alpha_n$ we get: 
\begin{equation}
 \mathbb{I} = -\frac{1}{\gamma_n} A^n - \frac{\gamma_1}{\gamma_n} A^{n-1} - \cdots - \frac{\gamma_{n-1}}{\gamma_n} A.   
\end{equation}
Multiplying all terms on the left by $A^{-1}$ gives: 
\[
A^{-1}\mathbb{I} = -\frac{\gamma_{n-1}}{\gamma_n}\mathbb{I} - \frac{\gamma_{n-2}}{\gamma_n}A - \cdots  -\frac{\gamma_1}{\gamma_n} A^{n-2} - \frac{1}{\gamma_n} A^{n-1}
\]
and post-multiplying by $b$ completes the proof that 
the solution lies in the Krylov basis. 
\begin{figure}
    \centering
    \begin{subfigure}[t]{0.45\textwidth}
        \centering
        \includegraphics[width=\linewidth]{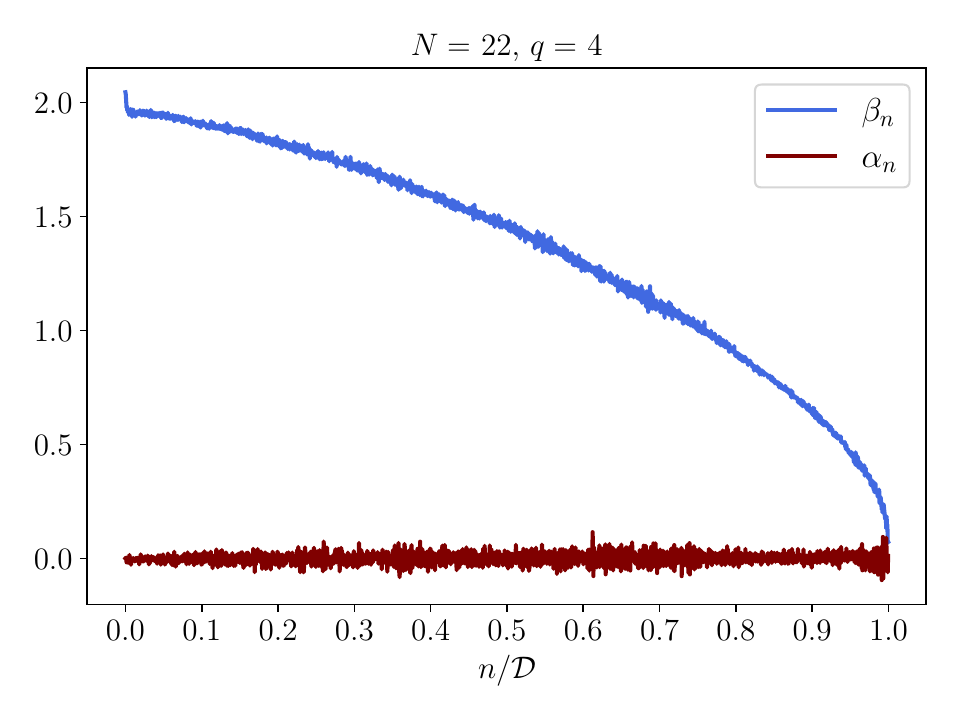} 
        \caption{Lanczos coefficients with orthogonalization with respect to all previous vectors with \code{gap=2}. See Algorithm~\ref{algo:algo1} for details.}\label{fig:yes}
    \end{subfigure}
    \hfill
    \begin{subfigure}[t]{0.45\textwidth}
        \centering
        \includegraphics[width=\linewidth]{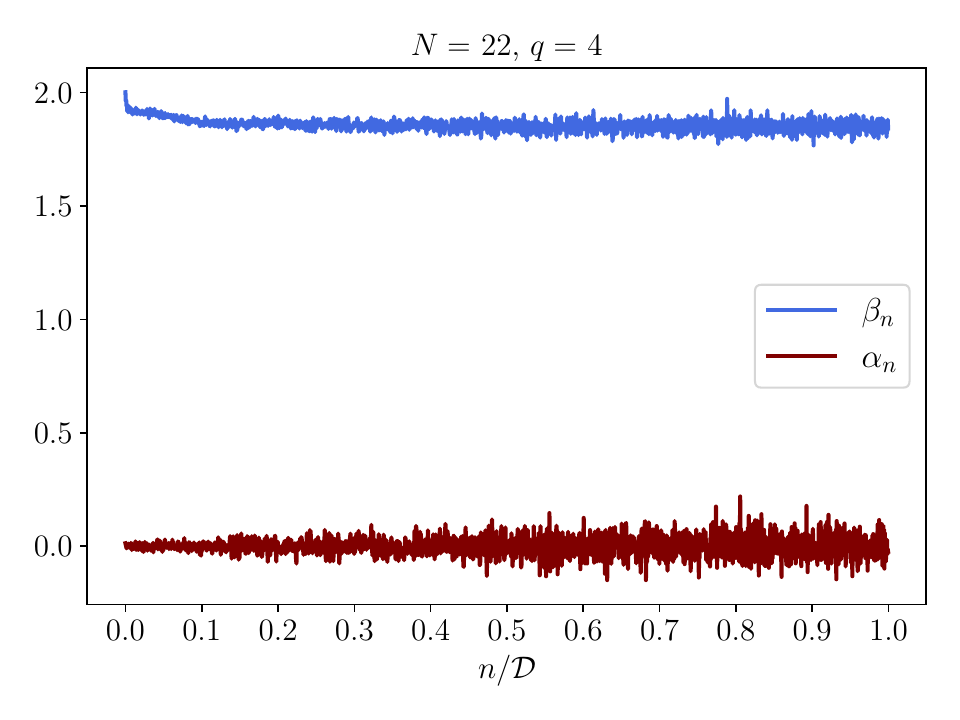} 
        \caption{Lanczos coefficients with no additional orthogonalization.} \label{fig:no}
    \end{subfigure}
    \caption{Disordered average Lanczos coefficients for $N=22$ sparse SYK with $k=1.6, q=4$. We find that it is necessary to orthogonalize every few iterations to ensure that spread complexity is reliably computed.}
\end{figure}

\begin{figure}[!h]
    \centering
    \begin{subfigure}[t]{0.48\textwidth}
        \centering
        \includegraphics[width=\linewidth]{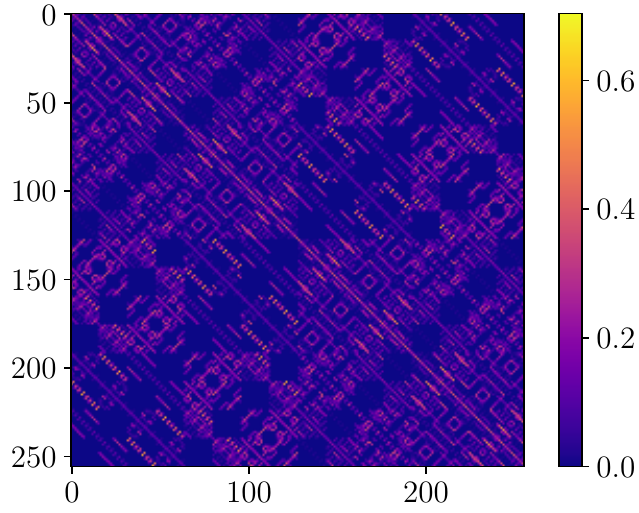} 
    \end{subfigure}
    \hfill
    \begin{subfigure}[t]{0.48\textwidth}
        \centering
        \includegraphics[width=\linewidth]{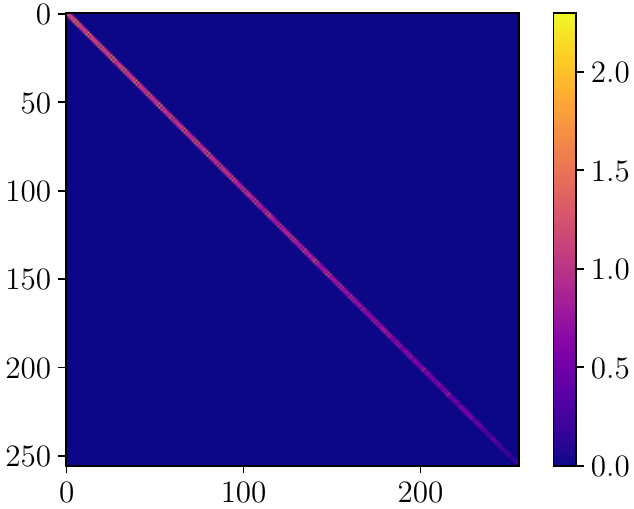} 
    \end{subfigure}
    \caption{\label{fig:cmap0}An example of $q=4$ sparse SYK Hamiltonian with $k=2$ for $N=16$ showing the absolute value of the elements. Left: The input Hamiltonian $H$. Right: After the Lanczos algorithm, we obtain the tridiagonal form where the diagonal elements are $\alpha$ while the `symmetric' off-diagonal elements are $\beta$.}
\end{figure}

The Gram–Schmidt procedure applied to $\left|\psi_n\right\rangle$ generates an ordered orthonormal basis $\mathfrak{K} = \{\left|K_n\right\rangle : n = 0, 1, 2, \cdots\}$ for the part of the Hilbert space explored by the time development of $\left|\psi(0)\right\rangle \equiv \left|K_0\right\rangle$. The major part of the numerical algorithm to compute the spread complexity is to solve for the Krylov basis using the Lanczos algorithm. Rather than writing the implementation of the algorithm from scratch, one can use the fact that the Hamiltonian is complex symmetric, and reducing it to Hessenberg form makes it tridiagonal. The sparse SYK Hamiltonian for $N=16$ and $k = 2, q= 4$ and the reduced Hessenberg form after implementation of the Lanczos algorithm are shown in Fig.~\ref{fig:cmap0}. So, in essence, by directly doing \code{hessenberg} which is available in \code{scipy}~\cite{scipy}, we can read off the Lanczos coefficients $\alpha$ and $\beta$. However, there are some caveats, as discussed in~\cite{Balasubramanian:2022tpr}. For this work, we explicitly follow the Lanczos algorithm and carry out full orthogonalization (FO) of every second vector to obtain the coefficients. The complete orthogonalization of a given vector with respect to all previous Krylov vectors is important to ensure that the computed peak value of complexity is accurate. We show a difference in the Lanczos coefficients especially $\beta$, with and without FO for $N=22$ 
in Fig.~\ref{fig:yes} and Fig.~\ref{fig:no}. However, FO adds considerably to the overall cost of the algorithm. Although there are other alternatives, such as selective orthogonalization~\cite{Parlett1979} and
partial orthogonalization~\cite{Simon1984}, 
we do not pursue them in this work. These improvements will be required to reach larger values of $N$. We provide the Lanczos algorithm\footnote{The algorithm is due to Cornelius Lanczos, a Hungarian mathematician and one of the Martians. The original motivation of the algorithm was the tri-diagonalization of a Hermitian matrix} in Algorithm~\ref{algo:algo1}. 
\begin{algorithm}
    \label{lanczsos}
    \begin{algorithmic}[1]
    \setstretch{1.3}
        \State $\ket{u_0}=\kp$ (Assume normalized)
        \State $\ket{x_1}= H \kp$
        \State $\a_1= \braket{x_1 \vert u_0} = \langle \psi \vert H \vert \psi \rangle $ 
        \State $\ket{v_1}=\ket{x_1}-\a_1\ket{u_0}$
        \For {$j = 1,2,\cdots N$}
            \State $\beta_j=\sqrt{\langle v_j \vert v_j \rangle}$
            \If{$\beta_j > 10^{-16}$}
                \State $\ket{u_j}\gets\frac{1}{\beta_j}\ket{v_j}$.
                \Else \State \code{break}
            \EndIf
            \State $\ket{x_{j+1}}= H\ket{u_j}$
            \State $\a_{j+1}= \braket{x_{j+1} \vert u_j}$
            \State $\ket{v_{j+1}}=\ket{x_{j+1}}-\a_{j+1}\ket{u_j}-\beta_j\ket{u_{j-1}}
            = H\ket{u_j} -\braket{x_{j+1} \vert u_j}\ket{u_j}-\sqrt{\langle v_j \vert v_j \rangle}\ket{u_{j-1}}$
            \If{$ \code{orthogonalization = True}$ and \code{j mod gap = 0}}
                \State $ \ket{v_{j+1}} = \ket{v_{j+1}} - \sum_{i=1}^{j} \langle u_{i} \vert v_{j+1} \rangle \ket{u_i}$
            \EndIf
        \EndFor
    \end{algorithmic}
\caption{\label{algo:algo1}Lanczos Algorithm. We input the starting state $\kp$ and Hamiltonian $H$ and the algorithm returns a set of $N$ orthonormal vectors $\{\ket{u_i}\}$ spanning the Krylov subspace and the coefficients $\alpha$ and $\beta$ known as Lanczos coefficients. In this paper, we do the \code{Line 16} only every second iteration. We refer to this as \code{gap}.}
\label{ref:algo1} 
\end{algorithm}
Once we have the coefficients $\alpha$ and $\beta$ at the end of the Lanczos algorithm, we use the first-order integrator to solve for $\phi_{n}(t)$ needed to compute the complexity defined in~\eqref{eq:def_KC}. We solve the differential equation as mentioned in Fig.~\ref{fig:cartoon1} using
\code{odeint} from
\code{scipy.integrate import odeint}. 

\bibliographystyle{utphys}
\raggedright
\bibliography{v3.bib}
\end{document}

%% file: styles.tex
\definecolor{halfgray}{gray}{0.55}
\definecolor{pyframe}{RGB}{207, 207, 207}
\definecolor{pybackground}{rgb}{0.95,0.95,0.95}
\definecolor{pypurple}{RGB}{170, 94, 255}

\definecolor{pykeyword}{HTML}{2A9D8F}
\definecolor{pyblue}{HTML}{4062BB}
\definecolor{ipython_red}{HTML}{52489C}
\definecolor{pycyan}{HTML}{E76F71}

\lstdefinelanguage{Python}{
    morekeywords={access,and,as,assert,break,class,continue,def,del,elif,else,except,exec,finally,for,from,global,if,import,in,is,lambda,not,or,pass,print,raise,return,try,while},
    morekeywords=[2]{True,False,self},
    sensitive=true,
    morecomment=[l]\#,
    morestring=[b]',
    morestring=[b]",    
    morestring=[s]{'''}{'''},
    morestring=[s]{"""}{"""},
    morestring=[s]{r'}{'},
    morestring=[s]{r"}{"},
    morestring=[s]{r'''}{'''},
    morestring=[s]{r"""}{"""},
    morestring=[s]{u'}{'},
    morestring=[s]{u"}{"},
    morestring=[s]{u'''}{'''},
    morestring=[s]{u"""}{"""},
    literate=
    {á}{{\'a}}1 {é}{{\'e}}1 {í}{{\'i}}1 {ó}{{\'o}}1 {ú}{{\'u}}1
    {Á}{{\'A}}1 {É}{{\'E}}1 {Í}{{\'I}}1 {Ó}{{\'O}}1 {Ú}{{\'U}}1
    {à}{{\`a}}1 {è}{{\`e}}1 {ì}{{\`i}}1 {ò}{{\`o}}1 {ù}{{\`u}}1
    {À}{{\`A}}1 {È}{{\'E}}1 {Ì}{{\`I}}1 {Ò}{{\`O}}1 {Ù}{{\`U}}1
    {ä}{{\"a}}1 {ë}{{\"e}}1 {ï}{{\"i}}1 {ö}{{\"o}}1 {ü}{{\"u}}1
    {Ä}{{\"A}}1 {Ë}{{\"E}}1 {Ï}{{\"I}}1 {Ö}{{\"O}}1 {Ü}{{\"U}}1
    {â}{{\^a}}1 {ê}{{\^e}}1 {î}{{\^i}}1 {ô}{{\^o}}1 {û}{{\^u}}1
    {Â}{{\^A}}1 {Ê}{{\^E}}1 {Î}{{\^I}}1 {Ô}{{\^O}}1 {Û}{{\^U}}1
    {œ}{{\oe}}1 {Œ}{{\OE}}1 {æ}{{\ae}}1 {Æ}{{\AE}}1 {ß}{{\ss}}1
    {ç}{{\c c}}1 {Ç}{{\c C}}1 {ø}{{\o}}1 {å}{{\r a}}1 {Å}{{\r A}}1
    {€}{{\EUR}}1 {£}{{\pounds}}1
    {^}{{{\color{pypurple}\^{}}}}1
    {=}{{{\color{pypurple}=}}}1
    {+}{{{\color{pypurple}+}}}1
    {*}{{{\color{pypurple}$^\ast$}}}1
    {/}{{{\color{pypurple}/}}}1
    {+=}{{{+=}}}1
    {-=}{{{-=}}}1
    {*=}{{{$^\ast$=}}}1
    {/=}{{{/=}}}1,
    literate=
    *{-}{{{\color{pypurple}-}}}1
     {?}{{{\color{pypurple}?}}}1,
    identifierstyle=\color{black}\ttfamily,
    commentstyle=\color{pycyan}\ttfamily,
    stringstyle=\color{ipython_red}\ttfamily,
    showstringspaces = false,
    keepspaces=true,
    showspaces=false,
    showstringspaces=false,
    rulecolor=\color{pyframe},
    frame=single,
    frameround={t}{t}{t}{t},
    framexleftmargin=6mm,
    numbers=left,
    numberstyle=\tiny\color{halfgray},
    backgroundcolor=\color{pybackground},
    basicstyle=\ttfamily\smaller[0.5],
    keywordstyle=\color{pykeyword}\ttfamily,
    keywords=[3]{EinsumPlanner, contract, einsum, ee, svd},
    keywordstyle={[3]\color{pyblue}},
}